\documentclass[journal]{IEEEtran}
\ifCLASSINFOpdf
\else
\fi
\ifCLASSOPTIONcompsoc
 \usepackage[caption=false,font=normalsize,labelfont=sf,textfont=sf]{subfig}
\else
 \usepackage[caption=false,font=footnotesize]{subfig}
\fi
\usepackage{url} 
\usepackage{graphicx}  
\usepackage{float}  
\usepackage{siunitx}
\usepackage{bm}
\usepackage{subfig}
\usepackage{makecell}
\usepackage{multirow}
\usepackage{algorithm}
\usepackage{algorithmic}

\usepackage{colortbl}
\usepackage{cite}
\usepackage{threeparttable}
\usepackage{amsfonts,amssymb} 
\usepackage{amsmath} 
\usepackage{amssymb}  

\begin{document}
%
\title{Multi-Variant Consistency based Self-supervised Learning for Robust Automatic Speech Recognition}

%
%
%

\author{Changfeng~Gao, ~\IEEEmembership{Student Member,~IEEE,}
        Gaofeng~Cheng, ~\IEEEmembership{Member,~IEEE,}
        Pengyuan~Zhang ~\IEEEmembership{Member,~IEEE}
\thanks{The authors are with the Key Laboratory of Speech Acoustics and Content Understanding, Institute of Acoustics, Chinese Academy of Sciences,
Beijing 100864, China. University of Chinese Academy of Sciences (e-mail: gaochangfeng@hccl.ioa.ac.cn;  chenggaofeng@hccl.ioa.ac.cn; zhangpengyuan@hccl.ioa.ac.cn).}
}

\maketitle

\begin{abstract}

Automatic speech recognition (ASR) has shown rapid advances in recent years but still degrades significantly in far-field and noisy environments. 
The recent development of self-supervised learning (SSL) technology can improve the ASR performance by pre-training the model with additional unlabeled speech and the SSL pre-trained model has achieved the state-of-the-art result on several speech benchmarks. 
Nevertheless, most of the previous SSL methods ignore the influence of the background noise or reverberation, which is crucial to deploying ASR systems in real-world speech applications.
This study addresses the robust ASR by introducing a multi-variant consistency (MVC) based SSL method that adapts to different environments. 
The MVC-SSL is a robust SSL pre-training method designed for noisy and distant-talking speech in real-world applications. 
Compared to the previous SSL method, the MVC-SSL can calculate the contrastive loss among audios from different acoustic conditions or channels and can learn invariant representations with the change in the environment or the recording equipment. 
We also explore different SSL training pipelines to balance the noisy distant-talking speech and extra high resource clean speech.
We evaluate the proposed method on the commercially-motivated dataset, CHiME-4, and the meeting dataset, AMI. With the help of the MVC-SSL and appropriate training pipeline, we can achieve up to 30\% relative word error rate reductions over the baseline wav2vec2.0, one of the most successful SSL methods for ASR.

\end{abstract}


%
\IEEEpeerreviewmaketitle

\section{Introduction}

Automatic speech recognition (ASR) is the primary technology for voice-based human-computer interaction.
Although the ASR models have realized promising performance on many tasks\cite{E2E-C-le-A,E2E-CTC-ATT,E2EASR1,E2EASR3,E2E-Online_transformer}, it still degrades significantly in far-field and noisy environments, especially when the training resource is limited. 
Recently, self-supervised learning (SSL) has shown excellent performance in the area of deep learning\cite{BERT,GPT-2,simclr}, including the ASR\cite{Wav2Vec,Wav2Vec2,conformer_fbank,hubert,wavlm,My_ICASSP}.
However, most of the previous SSL methods for ASR only focus on clean and close-talking speech like Librispeech\cite{libspeech} or Switchboard\cite{switchboard}, ignoring the influence of the background noise and reverberation. 
In this paper, we try to make the ASR model more reliable by designing a robust SSL method to face the background noise and reverberation in real-world scenarios like commercially-motivated, meeting, or conversation scenarios \cite{chime,chime5,ami,real-worldset-1,real-worldset-2}. 

Two problems need to be solved to design a robust SSL for the ASR. 
On the one hand, the robust SSL method should learn how to resist the various background noise and reverberation in the real-world environment. 
Because no matter what kind of training object is used, the principle of SSL is to leverage the data’s inherent co-occurrence relationships of the unlabeled speech \cite{qinghua-ssl}. 
Without the supervision of the label, the clean signal and the background noise play the same role in the self-supervised objective function.
If the signal-to-noise ratio (SNR) is low, the SSL model may pay more attention to the inessential background noise than the text-related speech signal. 
Some works try to improve the robustness of the SSL model by changing the model structure or training algorithm. 
\cite{wav2vec-c} proposes a wav2vec-C by combining a wav2vec2.0 \cite{Wav2Vec2} and VQ-VAE \cite{vq-vae} and evaluates on the far-field labeled data. 
\cite{wav2vec-r} trys to teach the model how to denoise by adding a reconstruction network during pre-training, which can also improve the noise robustness for the SSL. 
And \cite{avhubert} designs an AV-Hubert to improve robustness by complementing the audio stream with the visual information.

On the other hand, the robust SSL method should solve the domain-shifting between the clean speech and the noisy distant-talking speech because the training process of SSL needs extensive scale corpora. 
For example, \cite{Wav2Vec2,hubert,conformer_fbank} use more than 64 GPUs to pre-train their models on the LibriVox corpus, which contains about 60K hours of audiobook recordings. 
However, most of the public available real-world corpora\cite{chime, chime5, ami} only contain dozens of hours speech, due to the difficulty in collecting. 
So it is necessary to make use of the easily accessed reading or telephone corpora during pre-training and consider the mismatch between the reading or telephone corpora and the real-world corpora \cite{asr4real,asr4real2}.
Moreover, because of the heavy computation, many researchers have to seek the SSL model from model zoo \cite{transformers,torchhub} and then fine-tune it on different downstream task\cite{wav2vec-event,wav2vec-slu,wav2vec-spk,wav2vec-superb,deng1,deng2}, rather than pre-train a SSL model from scratch. But the public available SSL model may not fit the target domain. 
To solve the domain-shifting, \cite{Wav2Vec2-CL,xlsr,robustw2v} prove that by mixing up different data sources together, the SSL model can learn the knowledge from each data source.
\cite{robustw2v, continual-w2v} show that the continual-training is also a good method to utilize the high resource data. They pre-train the model on the high resource data and then transfer it to the limited resource data. 
However, these previous works mostly explore the domain-shifting between linguistic style, accent, or language. Slight background noise or reverberation could occur in their evaluation dataset.

In this study, we propose a novel multi-variant consistency (MVC) based SSL method to establish a robust ASR system. 
We first explore how to solve the domain-shifting between noisy distant-talking speech and additional clean reading speech during SSL pre-training. 
We choose the wav2vec2.0\cite{Wav2Vec2}, one of the most successful SSL methods for ASR, as the baseline and pre-train the model with different training pipelines, including source-data pre-training, data-mixing pre-training, and continual-training.
Results show that the continual-training is a suitable method to solve the domain-shifting.
After that, we use the proposed MVC-SSL pre-training to replace the baseline wav2vec2.0 pre-training. The proposed MVC-SSL can calculate the contrastive loss among audios from different channels or acoustic conditions. As a result, the MVC-SSL model can be more robust with the background noise and reverberation. 
We evaluate the proposed MVC-SSL on the CHiME-4 and AMI, two publicly available datasets for the commercially-motivated and meeting scenarios. 
Experiments show that the proposed MVC-SSL pre-training method can reduce 30\% relative word error rate (WER) over the baseline wav2vec2.0.



The rest of this paper is organized as follows: 
Section \ref{sec:related} introduces some related works.
Section \ref{sec:mvc-ssl} shows the details of the proposed MVC-SSL.
Section \ref{sec:pipeline} describes the training pipeline for the robust SSL.
Section \ref{sec:exp_set} and \ref{sec:exp-res} introduce the experiments and results.
Finally, Section \ref{sec:con} gives the conclusion.

\section{Related Works}\label{sec:related}

\subsection{Wav2vec2.0}\label{sec:w2v}

The wav2vec2.0 is one of the most successful SSL methods for ASR. The wav2vec2.0 model \cite{Wav2Vec2} contains a CNN-based feature extractor and a Transformer-based \cite{Transformer} context network. 
The CNN-based feature extractor consists of temporal convolution layers with GELU activation function. Each convolution layer has 512 channels with strides (5,2,2,2,2,2,2) and kernel widths (10,3,3,3,3,2,2). It will convert the input waveform $\bm{X}$ into compressed latent representation $\bm{Z}=\{\bm{z}_1,\bm{z}_2,...,\bm{z}_T\}$. And then, random mask will be applied on $\bm{Z}$. $p = 0.065$ of all time steps are chosen as the starting indices and then the next $M = 10$ consecutive time steps will be masked.
After masking, the Transformer-based context network will build contextualized representations $\bm{C}=\{\bm{c}_1,\bm{c}_2,...,\bm{c}_T\}$ from masked $\bm{Z}$ to capture high level content. In addition, a quantization module is used to convert $\bm{Z}$ into discrete representation $\bm{Q}=\{\bm{q}_1,\bm{q}_2,...,\bm{q}_T\}$.

The quantization module contains $G$ codebooks containing $V$ entries. And for each time step $t$, $\bm{z}_t$ is mapped to $\bm{l} \in \mathbb{R}^{G \times V}$ logits at first. Then the module use the Gumbel-Softmax \cite{gumbel-softmax} to choose one discrete entry $\bm{e}_g$ from the $g$-th codebook based on probability:
\begin{equation}\label{eq_q}
    p_{g,v}=\frac{\exp(l_{g,v}+n_v)/\tau }{\sum_{k=1}^{V}\exp(l_{g,k}+n_k)/\tau}
\end{equation}
where $\tau$ is a non-negative temperature, $g$ and $v$ are index of codebook and the entry. $n_k=-\log(-\log(u))$ and $u$ are uniform samples from 0 to 1. Finally, the discrete entry $\bm{e}_1,...,\bm{e}_G$ are concatenated and then transferred by a linear transformation to obtain the $\bm{q}_t$.

For the original wav2vec2.0 pre-training, the contrastive loss\cite{cpc,infonce} is calculated between the contextualized representations $\bm{C}$ and discrete representation $\bm{Q}$. For contextualized representation $\bm{c}_t$ at time step $t$, the positive sample is the discrete representation $\bm{q}_t$ at same time step, a negative samples set $\bm{Q}_t$ are chosen from other time steps, the contrastive loss is calculated as follow:

\begin{equation}\label{eq:cl}
    L_{c}=-\log{\frac{\exp(sim(\bm{c}_{t},\bm{q}_{t})/\kappa)}{\sum_{\widetilde{\bm{q}} \in \bm{Q}_{t}}\exp(sim(\bm{c}_{t},\widetilde{\bm{q}})/\kappa)}}
\end{equation}

\begin{equation}
    sim(\bm{a},\bm{b})=\frac{\bm{a}^T\bm{b}}{\left \| \bm{a} \right \|\left \| \bm{b} \right \|}
\end{equation}
$\kappa$ is the temperature factor and is set to 0.1. 

\begin{figure*}[tbp]
    \centering
    \includegraphics[width=13 cm]{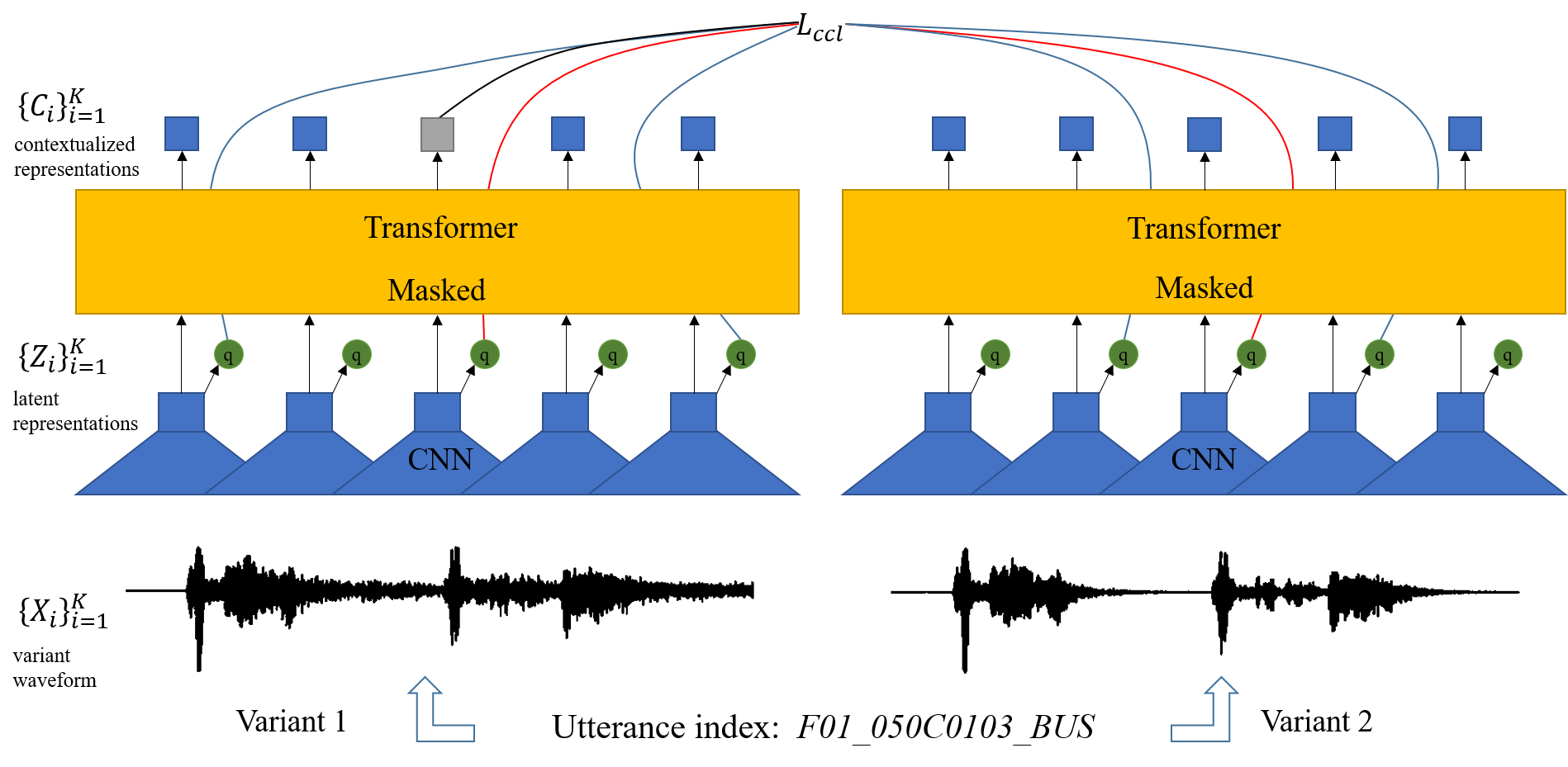}
    \caption{Computing method of the consistency contrastive loss. The two wav2vec2.0 models share the same parameters. The wav2vec2.0 model needs to distinguish the correct masked position in each variants waveform during pre-training. The black line stands for the anchor, the reds stand for positive samples, and the blues are negative samples. And the positive samples should be selected from the same time step.}
    \label{fig:model_struct}
\end{figure*}

\subsection{Domain-shifting for SSL}
Domain-shifting is a common problem for deep learning, and it can be more severe for SSL.
On the one hand, because of the heavy computation and the large-scale training set, most researchers need to seek SSL models from model zoos and research repositories rather than pre-train the model from scratch. However, the published pre-trained model is always pre-trained on the general domain rather than a task-specific domain. 
On the other hand, the task-specific domain data is much more difficult to collect than the general domain data and is not sufficient for the SSL. So it is necessary to make use of the easily accessed general-domain data during pre-training and solve the mismatch between the general data and the task-specific data.

For the domain-shifting, \cite{cv-ssl-transfer} designs a hierarchical pre-training pipeline that divides the SSL into source pre-training and target pre-training. 
\cite{cv-us-finetune} proposes the “unsupervised finetuning” method by adapting the representations from the source domain to the target domain. 
\cite{cv-ss-kd} boosts the SSL by transferring a task-standard SSL model into a task-specific SSL model via knowledge transfer.

In the ASR area, \cite{robustw2v} mixes up different domain speech to pre-train a wav2vec2.0 and finds that the model pre-trained on a variety of domains can perform well on each domain. However, the evaluation set in \cite{robustw2v} only includes clean speech with little reverberation.
\cite{Wav2Vec2-CL} also proves that the data-mixing method can be used in multi-lingual conditions.
Besides the data-mixing, \cite{continual-w2v} shows that the continual-training is also a good method to solve the domain shifting problem for the SSL.
However, whether these methods can be used in noisy and distant-talking speech needs to be explored.

\section{Multi-Variant Consistency SSL}\label{sec:mvc-ssl}

In this section, we describe the proposed MVC-SSL methods. First, the MVC-SSL uses various methods to simulate audios with different acoustic condition. Then, the MVC-SSL optimize the model with consistency contrastive loss to replace the contrastive loss in the wav2vec2.0 pre-training. Compared to the contrastive loss, the consistency contrastive loss can help the model maximize the representation measurement between the audios with different acoustic condition. So the MVC-SSL pre-trained model can be more robust than the baseline wav2vec2.0 pre-trained model.

\subsection{Consistency contrastive loss}

A robust ASR model should perform well with different background noise, reverberation, or different types of recording equipment.
For example, in the CHiME-4 challenge, the original audios are replayed and recorded in different environments or recorded by microphones at different positions. 
A human, even who does not know the language of speech, can tolerate these perturbations or distortions.
Therefore, a robust SSL model should also learn to convert the audios with different perturbations into invariant representations. 
Unfortunately, the current SSL method, including wav2vec2.0, cannot guarantee this consistency. 
Because during pre-training, they are only concerned with the relationship within one utterance. 
More seriously, as the speech is unlabeled, the clean signal and the background noise have equal contributions to the objective function. So when the SNR is low, the SSL model may learn the statistical characteristics of the background noise rather than the speech.	

To solve this problem, we designed a consistency contrastive loss to replace the traditional contrastive loss used in the wav2vec2.0.
The proposed consistency contrastive loss aims to provide consistency between the variants from the same original audio. The computing method of the consistency contrastive loss is shown in Fig \ref{fig:model_struct}.

Given audio $\bm{X}$, we apply different perturbations or distortions on it to construct a variants set $\{\bm{X}_i\}_{i=1}^K$. $K$ is the set size. Each $\bm{X}_i$ comes from the same origin but with different acoustic conditions or perturbations. 
Then, we feed $\{\bm{X}_i\}_{i=1}^K$ into the model to calculate the $\{\bm{Z}_i\}_{i=1}^K$, $\{\bm{C}_i\}_{i=1}^K$, $\{\bm{Q}_i\}_{i=1}^K$. Additionally, when mask the latent representation  $\{\bm{Z}_i\}_{i=1}^K$, same time steps should be chosen for each $\bm{Z}_i$. This operation aims to guarantee the masks are synchronous in every variants.

After getting $\{\bm{C}_i\}_{i=1}^K$, $\{\bm{Q}_i\}_{i=1}^K$, the consistency contrastive loss is calculated not only within one variant but also across variants from the same original audio. 
For each masked time step $t$ in the $i$-th variant audio, the model needs to identify the correct quantized latent speech representation $\bm{q}_{j,t}$ in each variant audios from the negative samples set $\bm{Q}_t$. The loss function is defined by the InfoNCE \cite{infonce} loss as follow:

\begin{equation}
        L_{ccl}=-\sum_{i,j}\log{\frac{\exp(sim(\bm{c}_{i,t},\bm{q}_{j,t})/\kappa)}{\sum_{\widetilde{\bm{q}} \in \bm{Q}_{t}}\exp(sim(\bm{c}_{i,t},\widetilde{\bm{q}})/\kappa)}}
\end{equation}
$i,j$ stand for the index of the variant.

According to  whether $i$ is equal to $j$, the consistency contrastive loss can be divided into self contrastive loss and cross contrastive loss as follow:
\begin{equation}\label{eq:ccl_split}
        L_{ccl}=L_{self} + L_{cross} 
\end{equation}
\begin{equation}\label{eq:ccl_self}
        L_{self}=-\sum_{i}\log{\frac{\exp(sim(\bm{c}_{i,t},\bm{q}_{i,t})/\kappa)}{\sum_{\widetilde{\bm{q}} \in \bm{q}_{t}}\exp(sim(\bm{c}_{i,t},\widetilde{\bm{q}})/\kappa)}}
\end{equation}
\begin{equation}\label{eq:ccl_cross}
        L_{cross}=-\sum_{i\ne j}\log{\frac{\exp(sim(\bm{c}_{i,t},\bm{q}_{j,t})/\kappa)}{\sum_{\widetilde{\bm{q}} \in \bm{Q}_{t}}\exp(sim(\bm{c}_{i,t},\widetilde{\bm{q}})/\kappa)}}
\end{equation}

\begin{figure*}[tbp]
    \centering
    \includegraphics[width=13 cm]{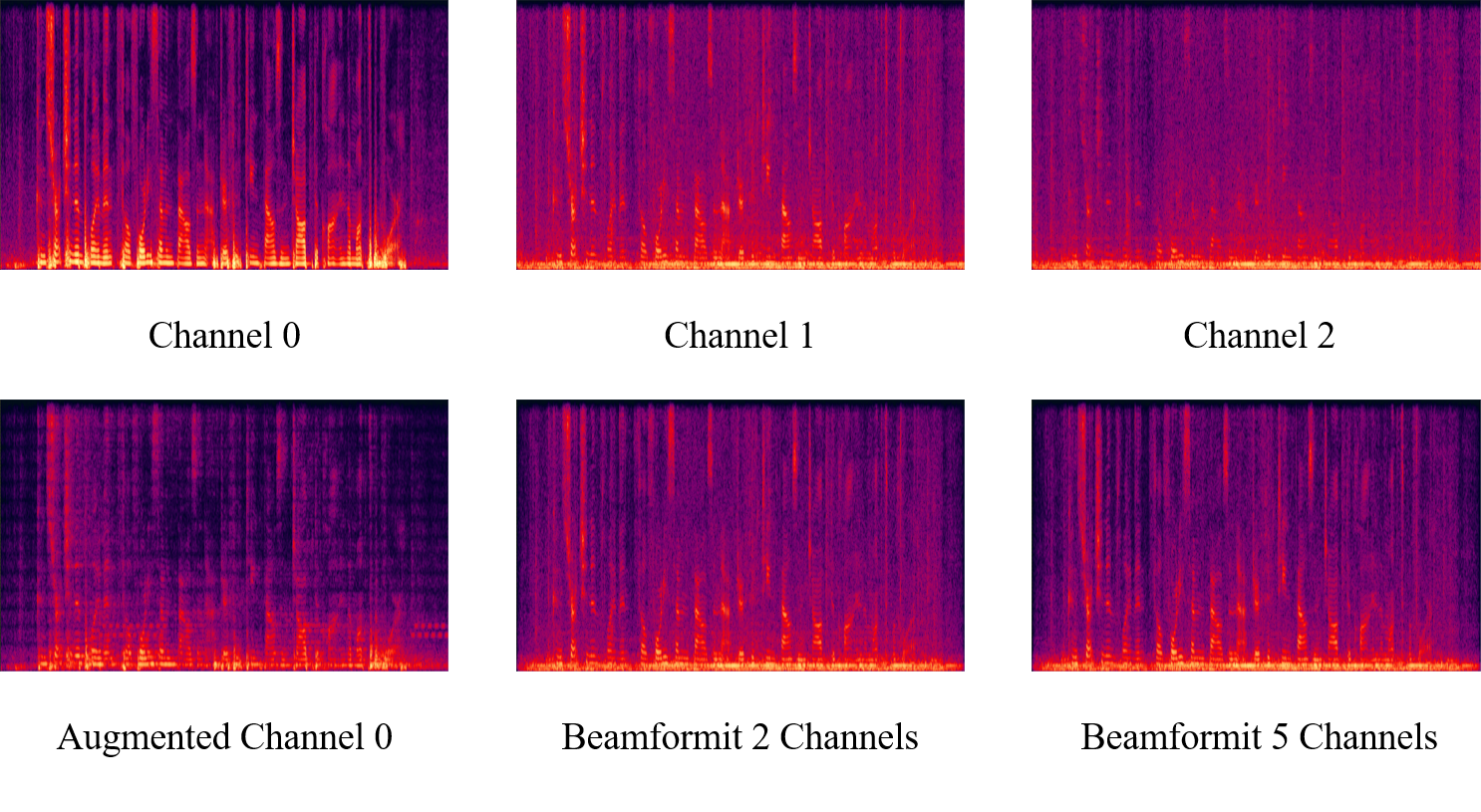}
    \caption{An example of the MVC. All above spectrograms are from the variants of the utterance \textit{F01\_050C0103\_BUS} in CHiME-4 corpus. 
    The audios from channel 0 is recorded by close-talking microphone and the channel 1 and 2 are from distant microphone. Audios in bottom line are augmented speech and enhanced speech. 
    For example, the consistency contrastive loss can be calculated between Channel 0 and Augmented Channel 0 (DA-MVC), Channel 1 and Channel 2 (MC-MVC), Channel 1 and Beamformit 2 (EH-MVC) or other pairs. 
    According to the consistency contrastive loss, the MVC-SSL model should map all of the variants into similar vectors in the representation space.
    As results, the pre-trained model can be more robust than the baseline wav2vec2.0, which treat these audios individually.}
    \label{fig:audio_version}
\end{figure*}

We can find that the consistency contrastive loss can be regarded as an expedition of the contrastive loss. The $L_{self}$ has the same format to the contrastive loss in Eq. (\ref{eq:cl}). 
However, the negative samples set $\bm{Q}_t$ can be sampled from all of the variant audios, and the traditional contrastive loss can only sample negatives from the original audio and other unrelated audios. This means that $L_{self}$ can provide more deceptive distractors. The $L_{cross}$ is an added part compared to the contrastive loss. Optimizing the $L_{cross}$ can maximize the measurement of the representations from different variant waveforms at the same time step. This means that the $L_{cross}$  can provide consistency between the audios from one origin but with different acoustic conditions. In other words, the SSL model can learn an invariant representation method under noise, reverberation, or other perturbation.

\subsection{Variants Set Construction}

The design of the variants set $\{\bm{X}_i\}_{i=1}^K$ is one of the key components of our proposed consistency contrastive loss. 
A well-designed variants set should simulate the original audio in different environments to help the model become robust to the complex acoustic condition.
Fortunately, there are sufficient methods to perturb the original waveform without changing the speech content, for example, recording the audios in different environments or simulating by different algorithms. 
So the consistency contrastive loss can be calculated from more than one angle. 
We call this phenomenon as MVC. An example of MVC is shown in Fig. \ref{fig:audio_version}.

\subsubsection{Data Augmentation (DA) Consistency}

We can create variants by applying data augmentation to the original audio to calculate the consistency contrastive loss. 
The augmentation should not change the audio tempo to guarantee that the augmented audios can align with the origin during contrast. 
We choose pitch shifting, reverberation, and adding background noise as the augmentation methods. 
For the pitch shifting, we randomly increase or decrease the pitch of the input audio. We randomly sampled the semitones number from -3 to 3. 
For the reverberation, we simulate the distant-talking speech by convoluting the audio with recorded room impulse responses (RIRs).
Moreover, for the adding background noise,  we mix the original audios with the real noise recordings and uniformly sample the SNR from 10dB to 30dB. 
We combine these data augmentation methods and apply them on-the-fly. The pitch shifting is applied with 50\% probability while others are applied with 15\% probability. So if none of them is applied to the audio, the audio will remain unchanged.

\subsubsection{Multi-Channel (MC) Consistency }
For distant-talking speech, the audios are always recorded by a microphone array containing more than one channel.
The previous SSL method cannot use the relationship between audios from different microphones, the essential information for multi-channel speech. 
In this study, we can use the channel information by regarding the audios from different channels as different variants from one origin audio. 
\footnote{Audios from different channels are not synchronous because of the position of the microphone. However, the offset is much smaller than the frame length (20ms) of the wav2vec2.0 so it can be ignored. }
As a result, consistency contrastive loss can help the model learn the commonality between audios from different channels and distinguish the difference caused by the positions of the microphones.


\subsubsection{Enhancement (EH) Consistency }

We can also calculate the consistency contrastive loss between noisy speech and enhanced speech. 
We randomly choose 2 or 5 microphones and then enhance them with a conventional delay and sum beamformer. The consistency contrastive loss will be calculated between the isolate channel speech and the enhanced speech.
We expect this consistency to help the SSL model learn how to denoise during pre-training.




\section{SSL Training Pipelines for Robust ASR}\label{sec:pipeline}

\begin{figure}[tbp]
    \centering
    \includegraphics[width=8 cm]{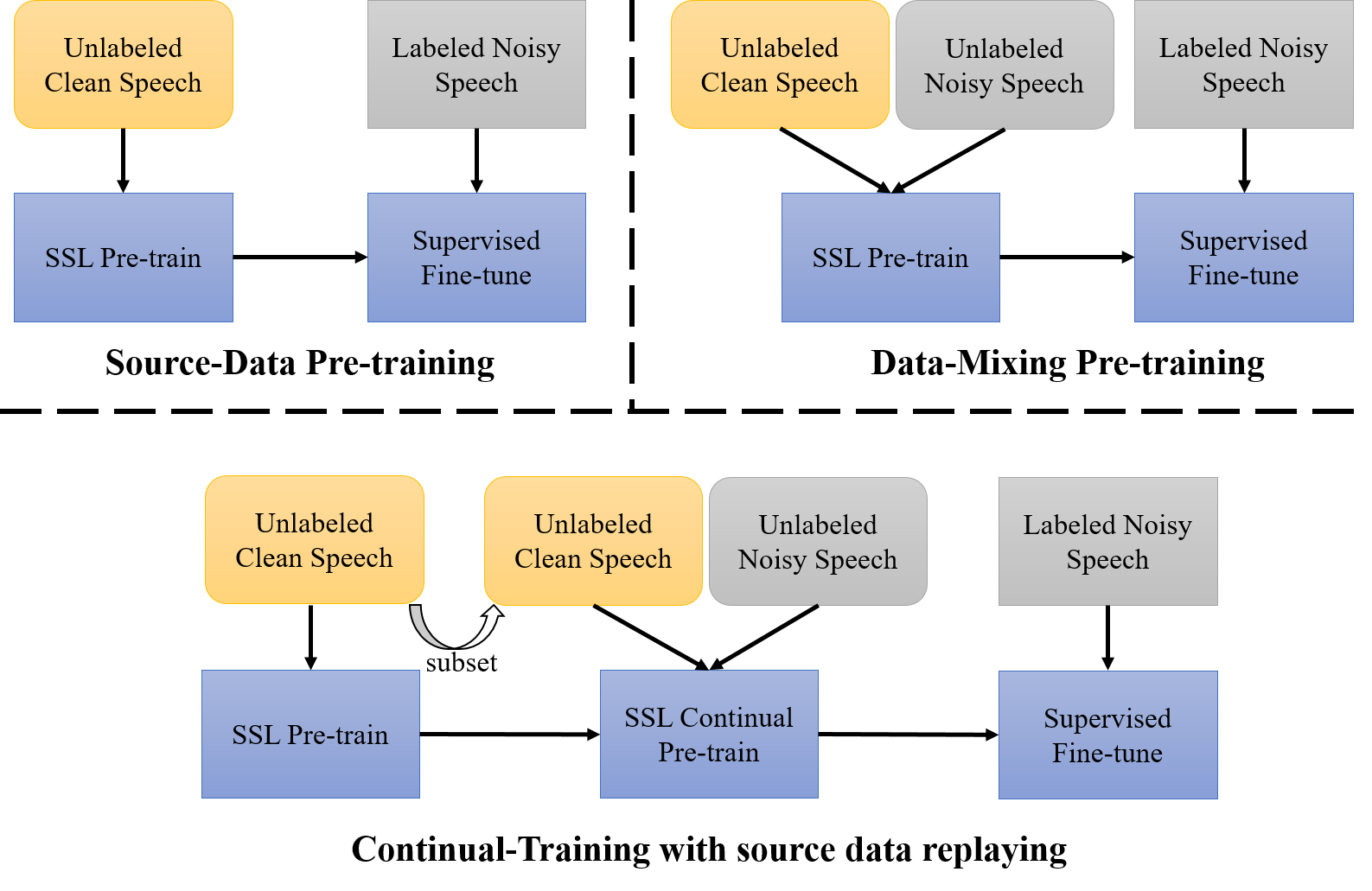}
    \caption{Different SSL training pipelines for robust ASR. The key points of the pipelines are how to balance the large scale clean speech and the limited noisy speech.}
    \label{fig:three_pretrain_method}
\end{figure}

The SSL pre-training corpus should be more than thousands of hours and even up to hundreds of thousand hours.
Besides the real-world corpora, we have to make use of the additional large-scale datasets \cite{fisher,gigaspeech,commonvoice,liblight}. And the large-scale datasets are always collected from audio-books, telephone, podcasts, or online videos rather than recorded from the real-world scenario. 
So it is crucial to design a training pipeline that can solve the domain-shifting problem between the small scale real-world corpora with noisy and distant-talking speech ($D_r$), and the much larger scale corpora with clean and close-talking speech ($D_c$). 
There are three training pipelines can make use of the $D_r$ and $D_c$, including source-data pre-training (Source-Data), data-mixing pre-training (Data-Mixing), and continual-training (Continual-Training) with source data replaying. 
Fig. \ref{fig:three_pretrain_method} shows the difference between these pipelines.
These pipelines have been used for many areas but not been sufficient explored in the robust SSL ASR task.
\subsubsection{Source-Data pre-training}
The Source-Data can be regarded as a baseline training pipeline for the SSL system, which only pre-trains the ASR model on the  $D_c$. 
This pipeline has been widely used in lots of downstream tasks\cite{wav2vec-event,wav2vec-slu,wav2vec-spk,wav2vec-superb,deng1,deng2} because most of the commonly available SSL models are pre-trained on the $D_c$.
Researchers can fine-tune the SSL model on the labeled speech directly with little computing power and realize promising performance on several tasks.

\subsubsection{Data-Mixing pre-training}
For the Data-Mixing, we put the speech in $D_r$ and $D_c$ together to pre-train the model like most of the previous works \cite{robustw2v, Wav2Vec2-CL, xlsr}. 
In this study,  the gap between $D_r$ and $D_c$ are more prominent because of the background noise, reverberation, and recording equipment. 
As the model should be pre-trained from scratch, this pipeline needs numerous computations.

\subsubsection{Continual-Training with source data replaying}
For the Continual-Training, we first pre-train the model on the $D_c$ and then continually pre-train it on the $D_r$. 
Additionally, we can also sample a subset $D'_c$ from $D_c$ then continually train the model on the $D'_c \cup D_r$. This operation is called source data replaying, and the rate between the size of the $D_r$ and $D'_c$ is the replaying rate. We believe that the replayed source speech can alleviate the catastrophic forgetting problem.
If the $D_c$ pre-trained model can be obtained from the model zoo, the computational intensity of this method will not be as great as the Data-Mixing.

\section{Experiment Setup}\label{sec:exp_set}

\subsection{Datasets}

We evaluate the SSL method on the recognition tasks of the publicly-available CHiME-4 corpus, for the commercially-motivated scenario, and AMI, for the meeting scenario. We choose these two corpora because they are close to practical application scenarios.
For the additional unlabeled clean and close-talking corpus, we use the audios from Librispeech\cite{libspeech} without transcription. 
Additionally, we also use real noise recording and RIRs for data augmentation.
Details of the corpora are described as follows:

\subsubsection{CHiME-4 Corpus}

The CHiME-4\cite{chime4} corpus focuses on the commercially-motivated scenario in the real-world application: a person talking to a mobile tablet device in real, noisy public environments, including streets, pedestrian areas, cafeterias, and buses.
The training data consists of 6-microphone recordings from the WSJ0 \cite{wsj0} corpus spoken live in the environments or simulated by mixing with noise backgrounds. 
The reverberation is not severe in the CHiME-4 corpus, but the SNR is very low (approximately 5 dB).
The total duration of the training corpus is only about 18 hours. 

\subsubsection{AMI Corpus}

The AMI\cite{ami} corpus contains approximately 78 hours of meeting recordings with three to five participants. The recording contains audio from independent headset microphones (IHM) worn by each participant and multiple distant microphones (MDM). Only using the first channel of the MDM for evaluation is referred to single distant microphone (SDM) setting.
Compared to the CHiME-4, the background noise in the AMI corpus is not too loud, but the reverberation is much more obvious.

\subsubsection{Librispeech Corpus}
Librispeech\cite{libspeech} is one of the most popular ASR corpora for SSL. It contains 1000 hours of close distance reading speech. We only use the audios in the Librispeech training set without any transcription. 

\subsubsection{RIRs and Noises}
Our experiments use RIRs and noises to generate the simulated noisy and distant-talking speech. The simulated RIRs and noises are provided by \cite{noise}, which contains 843 noise recordings sampled from the MUSAN \cite{musan} corpus and 60k simulated RIRs data. 

\subsection{SSL Recognition System}

We choose the wav2vec2.0 base model as the baseline recognition system. The model contains 12 transformer blocks. For each block, the attention dimension is 768 with 8 attention heads, and the inner dimensions of the FFN layer are 3072. The training pipeline of the wav2vec2.0 can be divided into pre-training and fine-tuning. All of the training is based on the  \textit{fairseq} toolkits \cite{fairseq}, and most of the training configurations are the same to \cite{Wav2Vec2}.

\begin{table*}[!t]
    \renewcommand{\arraystretch}{1.1}
    \caption{WER (\%) for models with different SSL training pipelines on CHiME-4 corpus.}
    \centering
    \begin{tabular}{|l|l|c c c c|c c c c|c c c c|}
    \hline
    \multirow{2}{*}{Training Pipeline}  & \multirow{2}{*}{LM}  & \multicolumn{4}{c|}{mic 1} & \multicolumn{4}{c|}{mic 2} & \multicolumn{4}{c|}{mic 6} \\
    & & dt-s & dt-r & et-s & et-r & dt-s & dt-r & et-s & et-r & dt-s & dt-r & et-s & et-r\\
    \hline
    Source-Data & Trigram & 10.8 & 8.5 & 17.2 & 15.5 & 8.6 & 6.8 & 13.6 &12.1 & 6.2 & 5.5 & 10.2 & 9.2 \\
    Data-Mixing & Trigram & 12.8&10.3&19.2&16.5&10.5&8.6&15.3&13.6&8.5&7.2&12.4&10.6 \\
    Continual-Training    & Trigram & 9.3&7.6&16.8&13.7&7.8&6.4&13.2&11.1&6.3&5.4&10.1&8.6 \\
      + 1:1 data replay  & Trigram& \textbf{9.0}&\textbf{7.3}&\textbf{15.8}&\textbf{13.1}&\textbf{7.4}&6.1&\textbf{12.1}&\textbf{10.5}&5.9&\textbf{5.1}&\textbf{9.4}&\textbf{8.0} \\
      + 1:3 data replay  & Trigram& 9.5&7.4&15.9&13.7&\textbf{7.4}&\textbf{6.0}&12.4&10.9&\textbf{5.8}&5.2&\textbf{9.4}&\textbf{8.0} \\
      + 1:9 data replay  & Trigram& 9.8&7.8&16.3&13.9&7.7&6.2&12.5&11.0&6.0&5.2&9.7&8.3 \\
    \hline
    Source-Data  & Transformer & 7.7&5.8&13.3&11.3&5.8&4.4&9.9&8.5&4.2&3.4&7.1&6.2 \\
    Data-Mixing & Transformer & 9.4&7.0&15.0&12.1&7.4&5.7&11.6&9.5&5.7&4.6&8.8&7.2 \\ 
    Continual-Training    & Transformer & 6.7&5.0&13.2&10.1&5.2&4.1&9.9&7.7&3.8&3.3&7.2&5.6 \\
      + 1:1 data replay  & Transformer& \textbf{6.4}&4.8&12.3&\textbf{9.7}&5.1&3.9&9.1&\textbf{7.3}&3.8&3.2&6.6&5.4 \\
      + 1:3 data replay  & Transformer& 6.5&\textbf{4.7}&\textbf{12.1}&\textbf{9.7}&\textbf{4.9}&\textbf{3.5}&\textbf{8.9}&\textbf{7.3}&\textbf{3.6}&\textbf{3.0}&\textbf{6.3}&\textbf{5.1} \\
      + 1:9 data replay  & Transformer& 6.9&4.8&12.4&10.1&5.1&3.6&9.0&7.4&\textbf{3.6}&\textbf{3.0}&\textbf{6.3}&5.3 \\
    \hline
    \end{tabular}
    \label{tab:result_three_method}
\end{table*}

\begin{table}[!t]
	\renewcommand{\arraystretch}{1.1}
	\caption{Comparison between self-supervised transfer and supervised transfer on the CHiME-4 corpus}
	\centering
	\begin{tabular}{|l|cc|cc|cc|}
		\hline
		\multirow{2}{*}{Training Pipeline} & \multicolumn{2}{c|}{mic 1}& \multicolumn{2}{c|}{mic 2} & \multicolumn{2}{c|}{mic 6} \\
		&dt-r&et-r&dt-r&et-r&dt-r&et-r  \\
		\hline
		Source-Data &8.5&15.5&6.8&12.1&5.5&9.2 \\
		Continual-Training &7.6&13.7&6.4&11.1&5.4&8.6 \\
		 + 1:1 data replay  &\textbf{7.3}&\textbf{13.1}&6.1&\textbf{10.5}&5.1&8.0 \\
		\hline
		supervised transfer (100h) &7.7&14.1&6.0&11.0&4.9&8.3 \\
        supervised transfer (960h) &7.4&14.3&\textbf{5.5}&\textbf{10.5}&\textbf{4.4}&\textbf{7.5} \\
		\hline
		
	\end{tabular}
	\label{tab:result_vs_st}
	
\end{table}

\subsubsection{Pre-training}

For pre-training, we first pre-train the model with the vanilla wav2vec2.0 method with the three training pipelines as the baselines. 
For the Source-Data, we use the official wav2vec2.0 model provided by the \textit{fairseq}, which is pre-trained on the Librispeech for 400k steps. 
For the Data-Mixing, we mix the speech from the Librispeech and CHiME-4/AMI and then pre-train the wav2vec2.0 on the mixed corpus for 400k steps. We train the model with Adam optimizer, warming up the learning rate for the first 32k updates to a peak of \num{5d-3}. We optimize the model with 8 GPUs and set the update frequency to 8, which equaling to optimizing with 64 GPUs. Except for the training corpus, all training configurations are the same as the official model.
Finally, to evaluate the Continual-Training, we use the Source-Data pre-trained model as the seed model and continually pre-train it on the CHiME-4/AMI for 20k steps. The learning rate of the Continual-Training is fixed to \num{5d-5}.
When using source data replaying, we sample a subset $D'_c$ from the Librispeech and then mix it with the CHiME-4/AMI ($D_r$) to continually train the model. 
We also evaluate various replaying rates between $D_r$ and $D'_c$ to find the best value.

We further compare the baseline wav2vec2.0 pre-training and the proposed MVC-SSL pre-training. When using MVC-SSL, we use the consistency contrastive loss to replace the contrastive loss in the wav2vec2.0 and pre-train the model from scratch. 
To calculate the consistency contrastive loss, the size $K$ of the variant waveform source is 2, because larger $K$ will consume too much cuda memory. 
For the Librispeech data, we can use DA-MVC to calculate the consistency contrastive loss. For the CHiME-4/AMI corpus, we use the MC-MVC and EH-MVC to create the variants.

\subsubsection{Fine-tuning}
After pre-training, we fine-tune the  ASR model on all the individual channels in the CHiME-4 challenge or the SDM in the AMI. 
We add a randomly initialized output layer on top of the pre-trained encoder to predict the output tokens in an end-to-end (E2E) way. 
The output tokens set contains 29 characters and a word boundary token. 
The Connectionist Temporal Classification (CTC) loss is used to optimize the models. 

\subsection{Language Model and Decoding}

We consider both n-gram and Transformer LM during decoding. For the n-gram LM, we use the task-standard trigram for the CHiME-4 challenge and the trigram trained by Kaldi for the AMI. 
For the Transformer LM, following most of the previous works, we use the text from the WSJ\cite{wsj} corpus for CHiME-4, and Fisher \cite{fisher} corpus for AMI for training.
The Transformer LM contains 16 layers of Transformer blocks. For each block, the attention dimension is 512 with 8 attention heads, and the inner dimensions of the FFN layer are 2048. 
For all decoding configurations, the LM weight is set to 2.0, and the word insertion penalty is set to 0.0. 
Beam 500 for the n-gram LM and beam 50 for the Transformer LM are used during beam searching decoding.

\subsection{Frontend Enhancement System}

We add a front-end enhancement system to transform the multi-channel noisy input signal into a single-channel enhanced output signal when decoding the multi-channel speech. Without special mention, the conventional delay and sum beamformer \cite{delay-sum-beamformer} is used for the experiments. 

\section{Experiment Results} \label{sec:exp-res}

\begin{table*}[htbp]
    \renewcommand{\arraystretch}{1.1}
    \caption{WER (\%) for models with different MVC SSL on CHiME-4 corpus.}
    \centering
    \begin{threeparttable}
    \begin{tabular}{|l|l|l l l l|l l l l|l l l l|}
    \hline
    \multirow{2}{*}{Training Pipeline}  & \multirow{2}{*}{LM}  & \multicolumn{4}{c|}{mic 1} & \multicolumn{4}{c|}{mic 2} & \multicolumn{4}{c|}{mic 6} \\
    & & dt-s & dt-r & et-s & et-r & dt-s & dt-r & et-s & et-r & dt-s & dt-r & et-s & et-r\\
    \hline
    Source-Data & Trigram & 10.8 & 8.5 & 17.2 & 15.5 & 8.6 & 6.8 & 13.6 &12.1 & 6.2 & 5.5 & 10.2 & 9.2 \\
     + DA-SIMU & Trigram &10.5 & 8.9 & 16.8 & 15.5 & 8.2 & 7.0 & 13.1 &11.7 & 6.4 & 5.6 & 10.0 & 8.8 \\
     + \{DA\}-MVC & Trigram & \textbf{9.5}&\textbf{7.3}&\textbf{15.7}&\textbf{13.3}&\textbf{7.5}&\textbf{6.1}&\textbf{11.9}&\textbf{10.7}&\textbf{5.9}&\textbf{5.1}&\textbf{9.4}&\textbf{8.2} \\
    \hline
    Continual-Training   & Trigram & 9.0&7.3&15.8&13.1&7.4&6.1&12.1&10.5&5.9&5.1&9.4&8.0 \\
     + DA-SIMU  & Trigram & 9.0& 7.2& 15.5& 13.0& 7.2& 5.9& 12.5& 10.2& 5.9& 5.2& 9.5& 7.9 \\
     + \{DA\}-MVC  & Trigram & 8.7&6.5&15.2&12.0&7.1&5.6&12.0&9.7&5.7&4.8&9.1&7.8 \\
     + \{MC\}-MVC  & Trigram & 9.0&7.0&15.0&12.7&7.1&5.7&11.5&10.3&5.6&4.9&8.8&7.5 \\
     + \{DA,MC\}-MVC  & Trigram & \textbf{8.4}&\textbf{6.3}&\textbf{14.7}&\textbf{11.4}&\textbf{6.6}&\textbf{5.4}&\textbf{11.4}&\textbf{9.3}&\textbf{5.3}&\textbf{4.6}&\textbf{8.4}&\textbf{7.3} \\
     + \{DA,MC,EH\}-MVC  & Trigram & 8.8&6.5&15.2&12.2&7.2&5.5&12.2&9.9&5.7&4.8&9.1&7.6 \\
    \hline
    Source-Data & Transformer & 7.7&5.8&13.3&11.3&5.8&4.4&9.9&8.5&4.2&3.4&7.1&6.2 \\
     + DA-SIMU & Transformer & 7.3&5.5&12.9&10.9&5.1&4.1&9.5&7.8&3.9&3.2&6.8&5.6 \\
     + \{DA\}-MVC & Transformer & \textbf{6.6}&\textbf{4.5}&\textbf{11.9}&\textbf{9.3}&\textbf{5.0}&\textbf{3.7}&\textbf{8.7}&\textbf{7.0}&\textbf{3.6}&\textbf{2.9}&\textbf{6.2}&\textbf{5.3} \\
    \hline
    Continual-Training  & Transformer & 6.4&4.8&12.3&9.7&5.1&3.9&9.1&7.3&3.8&3.2&6.6&5.4 \\
     + DA-SIMU  & Transformer & 6.4& 4.5& 11.8& 9.2& 4.7& 3.7& 9.1& 6.9& 3.7& 3.0& 6.6& 5.1 \\
     + \{DA\}-MVC  & Transformer & 6.2&4.2&11.7&8.5&4.8&3.5&8.8&6.6&3.7&2.9&6.3&4.9 \\
     + \{MC\}-MVC  & Transformer & 6.2&4.5&11.6&9.0&4.7&3.4&8.4&6.9&3.5&2.8&5.7&4.8 \\
     + \{DA,MC\}-MVC  & Transformer & \textbf{5.8}&\textbf{3.7}&\textbf{11.0}&\textbf{7.7}&\textbf{4.5}&\textbf{3.3}&\textbf{7.9}&\textbf{5.9}&\textbf{3.4}&\textbf{2.6}&\textbf{5.4}&\textbf{4.2} \\
     + \{DA,MC,EH\}-MVC  & Transformer & 6.2&4.3&11.6&8.6&4.8&3.5&8.7&6.4&3.5&2.8&6.3&4.8 \\
    \hline
    \end{tabular}
    \begin{tablenotes}
    \footnotesize
    \item DA, MC, EH stand for Data Augmentation, Multi-Channel and Enhancement Consistency.
    \end{tablenotes}
    \end{threeparttable}
    \label{tab:result_MVC}
\end{table*}

\subsection{Results for Wav2vec2.0 with Different SSL pipelines}

We first compare the three proposed SSL training pipelines on the baseline wav2vec2.0.  
For the Continual-Training, we also test the proposed source data replaying method with different replaying rates $D_r:D_c$. The WER on the CHiME-4 are shown in Table \ref{tab:result_three_method}. $s$ and $r$ is short for the $simulation$ and the $real$ in the table. 

According to the table, Continual-Training achieves the best performance among all three pipelines. Compared with the Source-Data baseline, Continual-Training on the CHiME-4 corpus can bring about 11.6\% relative word error rate reduction (WERR) on the single-channel \textit{et-r} and 6.5\% WERR on the six channel \textit{et-r}. The improvement in the enhanced speech is less than the signal-channel speech. 
Furthermore, the proposed source data replaying can improve the Continual-Training, especially for the enhanced speech. The relative WERR can increase to 13.0\% on the six-channel \textit{et-r}. This means that the source data replaying is more effective in the enhanced speech.
And among the tested replaying rate, the 1:1 is the best choice for all single-channel experiments. However, the best replaying rate is different when it comes to the multi-channel condition, especially when decoding with Transformer LM. It seems that adding more clean speech can get better performance. 
This is reasonable because the beamformer can transfer the multi-channel audio into cleaner single-channel audio by suppressing the background noise and the reverberation. So the enhanced audios will have more in common with the speech from the $D_c$.
The above results prove that the source data replaying in the Continual-Training can help the model maintain the original structure of the feature space, which can prevent catastrophic forgetting. However, too much source speech will hurt the transfer efficiency.

Another interesting result is that the Data-Mixing is not a good choice in this experiment, even worse than the baseline Source-Data. 
This result is different from the previous works, which study the domain-shifting between linguist-style or language \cite{robustw2v, Wav2Vec2-CL,xlsr}. 
We give the discussion as follows. The basic pronunciation units can be similar between different linguist-style or languages. So the pre-trained model can share the discrete tokens across different domains in the previous works. 
When it comes to noisy and distant-talking speech, same pronunciation units could have different waveforms because of the background noise and the reverberation. The Data-Mxing pre-trained model could regard them as two individual ones during pre-training. 
This could mislead the downstream ASR task. In contrast, the Continual-Training pre-trained model only needs to transfer the representation of the pronunciation unit from the clean condition into the noisy condition rather than learn a new representation method.

We also compare the Continual-Training with the supervised transfer on the CHiME-4 corpus. For the supervised transfer, the model is pre-trained on the unlabeled Librispeech, fine-tuned on labeled Librispeech, and finally, continually fine-tuned on the CHiME-4. Compared to the Continual-Training transfer,  supervised transfer needs the label of the $D_c$. Table \ref{tab:result_vs_st} lists the WER results with trigram decoding.

According to the table, we can find that all transferred models can be better than the baseline Source-Data model. 
The Continual-Training can be better on the most results on the single-channel task, especially when using source data-replaying. 
Nevertheless, the supervised transfer turns the tide when it comes to the multi-channel condition, because the enhanced data have more common with clean speech. This experiment proves that the SSL transfer can show better robustness on the change of the acoustic condition, and the supervised transfer can realize better performance when the target is similar to the source. However, Continual-Training transfer only needs unlabeled source data, while the supervised transfer needs the label of the source speech.

\subsection{Results for MVC-SSL on CHiME-4}

In this section, we evaluate the proposed MVC-SSL method on the Source-Data and the Continual-Training. Data-Mixing is discarded because of poor performance. Moreover, for the Continual-Training, we fix the replaying rate to 1:1. 
We only apply the DA-MVC for the Source-Data because only $D_c$ can be available during pre-training. For a fair comparison, we also pre-train the wav2vec2.0 on the simulated noisy speech with the same augmentation method (DA-SIMU), still using the original contrastive loss. 
For the Continual-Training, we apply all of DA-MVC, MC-MVC, and EH-MVC. The DA-MVC is used when pre-training on the Librispeech speech. While the  MC-MVC and EH-MVC are used for the multi-channel CHiME-4 speech. Results are shown in Table \ref{tab:result_MVC}.

The table shows that the proposed MVC method can reduce the WER for both the  Source-Data and the 
Continual-Training.
For the Source-Data, we can find the DA-MVC can bring about 10\% performance improvement compared to the baseline and is also better than DA-SIMU.
This indicates that the improvement of the DA-MVC comes from two aspects. 
For the first aspect, the data augmentation can expand the training data and alleviate the overfitting during pre-training like the DA-SIMU. 
More importantly, the consistency contrastive loss can help the model learn a consistent representation method between augmented and clean audio. So the representation can be more robust than the DA-SIMU.

For the Continual-Training, we can find that the model pre-trained with all DA-MVC and MC-MVC can realize the best performance. On the signal-channel \textit{et-r}, it can bring about 20.6\% WERR than the Continual-Training model without MVC-SSL and 31.0\% WERR than the baseline Source-Data model.
Unlike our prediction, adding additional EH-MVC shows little benefit, whether in single-channel or multi-channel conditions. 
This result is consistent with the previous supervised robust ASR system\cite{chime}, which finds that training the systems on all noisy individual channels can be more effective than training on one single-enhanced signal.
We explain that the denoising operation will change the details of the waveform, which could be essential information for the SSL.  

\subsection{Results for MVC-SSL on AMI}

\begin{table}[t]
	\renewcommand{\arraystretch}{1.1}
	\caption{WER (\%) for models with different MVC SSL on AMI SDM corpus.}
	\centering
	\begin{threeparttable}
	\begin{tabular}{|l|cc|cc|}
		\hline
		\multirow{2}{*}{Training Pipeline} & \multicolumn{2}{c|}{Trigram} & \multicolumn{2}{c|}{Transformer} \\
		&dev&eval&dev&eval \\
		\hline
		Source-Data &29.8&32.9&29.2&32.5 \\
		 + \{DA\}-MVC &\textbf{28.2}&\textbf{31.4}&\textbf{26.9}&\textbf{30.2} \\
		\hline
		Continual-Training &29.2&32.5&28.0&31.3 \\
		 + \{DA\}-MVC &27.9&\textbf{31.1}&26.7&30.2 \\
		 + \{MC\}-MVC &28.8&32.3&27.7&31.2 \\
		 + \{DA,MC\}-MVC &\textbf{27.6}&31.3&\textbf{26.4}&\textbf{29.8} \\
		\hline
		
	\end{tabular}
	\end{threeparttable}
	\label{tab:result_ami}
\end{table}   

In the subsection, we apply the proposed MVC-SSL to the AMI corpus. We evaluate the MVC-SSL on the SDM track, the most challenging task in the AMI corpus. The results are shown in Table \ref{tab:result_ami}.

According to the table, we can find that our proposed MVC-SSL can also show excellent performance in the meeting scenario. On the eval set, about 7.1\% and 4.8\% relative WERR can be realized on the Source-Data and Continual-Training.
This means that the consistency contrastive can tell the model learn both denoising and de-reverberating. 

We can also find some differences from the results on the CHiME-4. 
When the MVC-SSL is used, the gap between the Source-Data and the Continual-Training is less significant than the CHiME-4. 
The Source-Data model and the Continual-Training  model can realize similar WER when using DA-MVC.
This means that the Source-Data model with DA-MVC can be robust enough on the reverberation of the AMI corpus, so adapting to the $D_r$ is not as necessary as the CHiME-4. 
Another difference is that compared to the CHiME-4, the DA-MVC shows more effect than the MC-MVC. 
Because during DA-MVC pre-training, we can convolute the audio with different RIRs. However, for the MC-MVC, all channels are recorded in one room. So the DA-MVC pre-trained model can learn the reverberation-independent representations, while the representations from the MC-MVC will be related to the information of the RIRs.
In other words, DA-MVC has more ability to de-reverberate than the MC-MVC.

\subsection{Comparing with Previous Works}

In this subsection, we compare our proposed MVC-SSL with the previous robust ASR system. To our acknowledgment, for most real-world environment tasks, the hybrid systems always show better performance than the E2E system \cite{mc-e2e,mc-e2e2}, especially with low resources.

For the CHiME-4 task, we compare our results with previous works on the single-channel task and the six-channel task. 
Moreover, for the six-channel task, we use a DNN-Beamformer\cite{dnn-bfeamformer} to replace the previous delay and sum beamformer to improve the performance further. Results are shown in Table \ref{tab:result_previous}.

\begin{table}[t]
    \renewcommand{\arraystretch}{1.1}
    \caption{WER (\%) comparisons with previous works on CHiME-4 corpus.}
    \centering
    \begin{tabular}{|l|l|cc|cc|}
    \hline
    \multirow{2}{*}{Method} & \multirow{2}{*}{Architecture} & \multicolumn{2}{c|}{mic 1} & \multicolumn{2}{c|}{mic 6} \\
     &  & dt-r & et-r & dt-r & et-r \\
    \hline
    Kaldi\cite{kaldi-chime} & hybrid-TDNN & 5.6 & 11.4 & 1.9 & 2.7 \\
    Du et al.\cite{ifly-chime} & hybrid-DCNN & 4.6 & 9.2 & 2.1 & 2.2 \\
    Wang et al.\cite{sota-chime} & hybrid-BLSTM & 3.5 & 6.8 & 1.5 & 2.0 \\
    \hline
    ESPNet egs\cite{espnet-conformer} & E2E-Conformer & 11.7 & 20.6 & 7.9 & 14.2 \\
    Guo et al.\cite{guo-chime} & E2E-Transformer & -- & -- & 15.8 & 26.8 \\
    Tsunoo et al.\cite{chime-tts} & E2E-Conformer & 8.4 & 15.7 & -- & -- \\
    Wang et al.\cite{wav2vec-r} & E2E-wav2vec2.0 & 5.0 & 9.0 & -- & -- \\
    MVC-SSL (Ours) & E2E-wav2vec2.0 & 3.7 & 7.7 & 2.0 & 3.5 \\
    \hline
    \end{tabular}
    \label{tab:result_previous}

\end{table}    

The table shows that our model can realize better performance than the previous E2E system on both single-channel and six-channel tracks. 
Our system realizes comparable results to the state-of-the-art hybrid ASR system on the single-channel track. 
The hybrid systems always use lots of technologies like lattice rescore, speaker adapter \cite{spkadp}, and information fusion \cite{fusion} while our system is much simpler. 
On the six-channel condition, hybrid systems show better performance than our system, especially for the real data. One of the reasons is that our frontend system performs better on the simulated speech than on the real recording. 
Nevertheless, our model can still be much better than the previous multi-channel E2E ASR systems, which combine the frontend enhancement and the backend recognition in one neural network \cite{guo-chime}.

We also compare our method with the previous works on the AMI SDM corpus. Most of the previous E2E systems are evaluated on the IHM or use additional transcribed speech because the performances of the E2E systems are always worse than the hybrid systems. Results are shown as in Table \ref{tab:result_previous_ami}.

\begin{table}[tbp]
	\renewcommand{\arraystretch}{1.1}
	\caption{WER (\%) comparisons with previous works on AMI corpus.}
	\centering
	\begin{threeparttable}
	\begin{tabular}{|l|l|cc|}
		\hline
		Method & Architecture & dev & eval \\
		\hline
		Kaldi \cite{ami-kaldi}& hybrid-BLSTM & 42.5 & 45.6 \\
		Peddinti et al. \cite{pedd-ami} & hybrid-TDNN-BLSTM & 37.0 & 40.4 \\
		Kanda et al. \cite{kanda-ami}& hybrid-CNN-TDNN-BLSTM & 33.4 & 36.4 \\ 
		\hline
		ESPNet egs\cite{espnet} \tnote{1} & E2E-Transformer & 30.4 & 32.0 \\
		Kanda et al. \cite{kanda-e2e-ami}& E2E-Conformer & 47.5 & 51.1 \\
		Kanda et al. \cite{kanda-e2e-ami} \tnote{2}& E2E-Conformer & 23.0 & 25.8 \\
		MVC-SSL (Ours) & E2E-wav2vec2.0 & 26.4& 29.8 \\
		\hline
	\end{tabular}
	\begin{tablenotes}
	\footnotesize
	\item[1] Results for IHM.
	\item[2] Using additional 75k hours of clean labeled speech.
	\end{tablenotes}
	\end{threeparttable}
	\label{tab:result_previous_ami}
	
\end{table}    

Our model can outperform most of the previous systems on the AMI corpus. Compared to the state-of-the-art system in \cite{kanda-e2e-ami}, we only use additional 1k hours of unlabeled speech in Librispeech for pre-training, while \cite{kanda-e2e-ami} need to pre-train their model on 75k hours of labeled speech. Without additional labeled data, the WER for \cite{kanda-e2e-ami} will increase to 51.1\% on the eval set.
This indicates that on the AMI corpus, the SSL-based pre-training also has the potential to realize comparable performance to the supervised pre-training, which uses paired speech and transcription. 

\section{Conclusion}\label{sec:con}
 
In this study, we propose a MVC-SSL pre-training method to improve the robustness of the ASR system. 
Compared with the previous SSL method, the MVC-SSL can learn an invariant representation with the change in the environment by the consistency contrastive loss.
We also explore three training pipelines to solve the domain-shifting problem between the noisy and distant-talking speech and the clean and close-talking speech, including Source-Data pre-training, Data-Mixing pre-training, and Continual-Training with source data replaying. 
We evaluate our methods on the CHiME-4 and AMI.
Experiments show that Continual-Training with source data replaying is the best method to take advantage of the additional clean speech.  
The proposed MVC-SSL can achieve up to 30\%  relative WERR than the baseline wav2vec2.0 and realize a comparable result to the state-of-the-art systems on CHiME-4 and AMI. 
This study proves that the proposed MVC-SSL is an effective method to solve the robust problem for the ASR system, one of the most crucial problems for real-world speech applications.

\section*{Acknowledgment}
This work is partially supported by the National Key Research and Development Program (No. 2020AAA0108002).

\ifCLASSOPTIONcaptionsoff
  \newpage
\fi



\normalsize
\bibliographystyle{IEEEtran}
\bibliography{main}

\end{document}